\documentstyle[prb,aps,multicol,epsfig]{revtex}
\newcommand{\braced}[1]{\langle#1\rangle}
\begin {document} 
\draft
\title{Vortex motion in superconducting 
YBa$_2$Cu$_3$O$_{7-\delta}$ inferred from the damping of the oscillations 
of a levitating magnetic microsphere.}
\author{R. Grosser, P. H\"ocherl, A. Martin, M. Niemetz, and W. Schoepe}
\address{Institut f\"ur Experimentelle und Angewandte Physik,
Universit\"at Regensburg, D-93040 Regensburg, Germany}
\date{\today}
\maketitle
\begin {abstract}
The damping of the oscillations of a small permanent magnet 
(spherical shape, radius 0.1 mm) levitating between two parallel  
YBCO surfaces is measured as a function of oscillation amplitude and
temperature. 
The losses in the samples (epitaxial thin films, bulk granular and bulk 
melt-textured) are analyzed in terms of oscillating shielding currents 
flowing 
through trapped flux lines whose motion gives rise to electric fields.
We find dissipation to originate from different mechanisms of flux dynamics. 
At small amplitudes there is a linear regime described by a surface 
resistance varying from $10^{-9}\,\Omega$ for the bulk samples down to 
$10^{-13}\,\Omega$ for the thin films at low temperatures.
With increasing amplitude various nonlinear regimes are observed, firstly 
collective pinning with diverging energy barriers, secondly in bulk 
samples above 85\,K hysteretic damping, 
and finally in thin films exponentially large losses which can be 
described by pinning energies vanishing linearly at large currents.

\end{abstract}
\begin{multicols}{2}  
\section{Introduction}
A permanent magnet levitating above, or suspending below, a 
superconducting surface has an equilibrium position which is determined by 
magnetic forces due to screening currents and trapped flux.\cite{rf1}
Except for a small logarithmic long-term relaxation of the 
magnetization of the superconductor this position is stable. The system is 
in a metastable state with stationary flux lines and consequently without 
any decay of the dc supercurrents, i.e., without any energy dissipation. 
However, oscillations of the levitating magnet about its equilibrium 
position give rise to ac magnetic fields at the surface of the 
superconductor which modulate the shielding currents.  Flux lines are  now 
moved back and forth periodically as a result of the oscillating component 
of the Lorentz force.
This leads to
energy dissipation which we can determine from measurements of the 
oscillation amplitude of the magnet as a function of an 
external driving force.
The measured losses can be attributed to various regimes of flux motion in 
the superconductor.

In our present work we have investigated the energy 
dissipation when the magnet is levitating between two horizontal 
YBa$_2$Cu$_3$O$_{7-\delta}$ (YBCO) surfaces either of bulk material (both 
sintered and melt-textured) and of epitaxial thin films. In all samples and 
at all temperatures we find that at small amplitudes of the ac shielding 
currents ($\simeq 10\,\mathrm{A/m}$) the dissipation can be described by 
a 
surface resistance which varies from $10^{-9}\,\Omega$ for the bulk samples 
down to below $10^{-12}\,\Omega$ for the thin films at 4\,K. At larger 
amplitudes nonlinear behavior is observed which can be attributed to 
thermally activated flux flow with pinning energies whose diverging current 
dependence is typical for collective pinning. In addition, for a given 
current, the pinning energies have a linear temperature dependence 
which is an indication of flux creep. Upon 
further increase of the drive and at 
temperatures above 85\,K we find that the dissipated power in the 
bulk samples varies with the cube of the amplitude. We discuss this regime 
in terms of hysteretic dissipation having a coefficient which diverges near 
$T_c$. In the films, however, a quite different behavior is 
observed 
instead, namely, an exponential increase of the dissipation which ultimately 
exceeds even the bulk values. In this regime we can describe the data by a 
Kim-Anderson type model of flux motion with linearly vanishing pinning 
energies at large currents. In summary, by varying the drive of the 
oscillating magnet by five orders of magnitude we can study different 
regimes of vortex dynamics in YBCO at a given temperature.
Some of our preliminary results on granular samples \cite{apl1} and on thin 
films \cite{xxx} have already been published in short form. In this 
article 
a more detailed analysis and a more complete account of our work is 
presented.

This article is organized as follows. In the next Chapter we describe our 
experimental method, and in Ch.III the model used for data analysis is 
outlined. In Ch.IV we present the results of our experiments for the thin 
films and for the bulk samples. Chapter\,V concludes our work.

\section{experimental method}
Our experimental method 
consists of placing a magnetic microsphere made of SmCo$_5$ (radius 
0.1\,mm, mass $m = 5.2\cdot 10^{-8}\,\mathrm{kg}$) 
between two horizontal surfaces (spacing 1\,mm, diameter 4\,mm) of a 
parallel-plate capacitor made of YBCO. This is an extension of earlier 
work, where the capacitor was made of niobium\cite{niob} and the levitating 
sphere was used for hydrodynamic studies in superfluid 
helium.\cite{helium} In Fig.~1 a schematic of our electronics is shown. 
Before the capacitor is cooled through $T_c$ we apply 
a dc voltage of about 800\,V to the bottom electrode. Therefore, the magnet 
carries an electric charge $q$ of about 2\,pC when levitating. The dc 
voltage is then switched off and oscillations of the magnet can be excited 
with an ac voltage $U_{ac}$ ranging typically from 0.1\,mV to 20\,V and 
having a frequency at the resonance of the oscillations ($\approx 
300$\,Hz). These vertical oscillations induce a current $qv/d$ in the 
electrodes, where $v$ is the velocity of the magnet, which is measured by 
an electrometer connected to a lock-in amplifier. The capacitively coupled 
signal due to the driving voltage $U_{ac}$ which is superimposed upon the 
signal from the oscillating magnet is electronically nulled off resonance 
by 
adding the inverted pickup of appropriate amplitude. For a given amplitude 
of the driving force $F = qU_{ac}/d$ we measure the 
maximum signal when slowly sweeping the frequency towards resonance.
Because at large oscillation amplitudes the restoring forces become 
nonlinear, the resonance curves are usually hysteretic and care has to be 
taken to achieve resonance. We therefore also monitor the phase of the 
signal with respect to that of the drive. In the Appendix a short 
discussion of some of the elastic properties of this oscillator
is presented. Further data are taken either by varying the driving force at 
constant temperature or by sweeping the temperature at a fixed driving 
force while keeping the oscillator at resonance by adjusting the frequency 
of the drive. 
\begin{figure}
\centering\epsfig{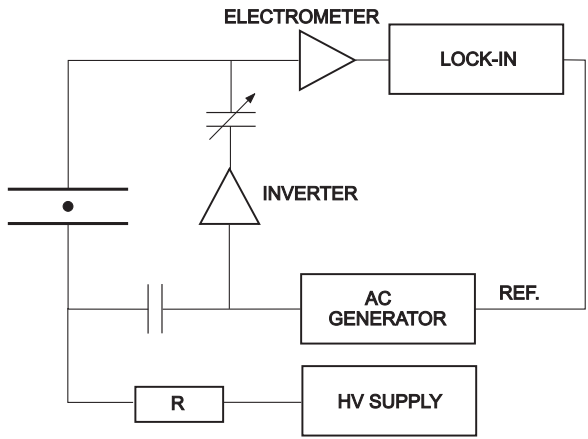}\\[1ex]
\begin{minipage}{8.6cm}\caption{Schematic of the experimental setup. The 
electrically charged magnetic microsphere is levitating inside the 
superconducting YBCO capacitor.
}\end{minipage}
\label{fig22}
\end{figure}%
In order to infer the 
velocity amplitude from the measured current amplitude and the driving force 
from the ac voltage we determine the charge $q$ by recording a resonance 
curve at very small drive (which is not hysteretic) and by analyzing its 
width 
(which is given by the damping coefficient $\gamma$) and its height which 
then 
determines the charge. Alternatively, in particular at low damping, 
we measure the time constant $\tau$ of the freely decaying oscillations
from which we obtain $\gamma=2m/\tau$. Comparing this result with the
driven response at the same temperature gives the charge.
The driving forces range from 10$^{-13}$\,N to 10$^{-7}$\,N, the velocity 
amplitudes from 0.1 mm/s to 50 mm/s, and the oscillation amplitudes $a = 
v/\omega$ from 50\,nm to 30\,$\mu$m. The dissipated power varies from 
10$^{-9}$\,W down to below 10$^{-17}$\,W. 

The YBCO we used for the superconducting capacitor was either home-made 
sintered material\cite{apl1} (thickness 2\,mm) or melt-textured 
samples\cite{mtg} of 
the same geometry, and  epitaxial thin films\cite{xxx} 
having a thickness of either 450\,nm or 190\,nm laser deposited on 
insulating substrates of SrTiO$_3$ (lower electrode) and 
Y$_2$O$_3$-stabilized ZrO$_2$ (upper electrode) 
both cut with an 
inclination of 2 degrees off the (001) orientation. The inclination 
improves both flux pinning and, due to initiating a terrace growth, 
crystallinity of the films. The lower electrode is protected by a 20 nm 
thick PrBa$_2$Cu$_3$O$_7$ epitaxial layer. The YBCO films have an 
extremely sharp superconducting transition at $T_c=89.5$~K measured by an 
ac susceptibility method, the complete transition width being 0.1 K. 
Further details of the thin film production can be found in 
Ref.\ \onlinecite{films}.

\section{Dipole Model}
The oscillations of the magnet give rise to ac surface currents which exert 
an oscillating Lorentz force on the flux lines. This causes energy 
dissipation which can be determined from the damping of the oscillations 
either by measuring the amplitude as a function of the driving force or by 
recording a free decay. In this section a brief discussion of a model 
used for a quantitative analysis of our data is given. 
\begin{figure}
\centering\epsfig{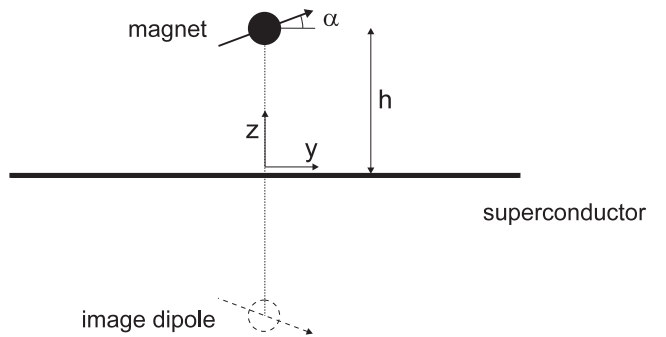}\\[1ex]
\begin{minipage}{8.6cm}\caption{Geometry of the dipole moment above the 
superconducting surface and its image below.}\end{minipage}
\label{fig42}
\end{figure}

In our experiment the superconductors are cooled in the field of the magnet
and therefore flux is trapped in the superconducting electrodes. We estimate 
the density of the trapped flux to be $n\Phi_0 \sim 1$\,mT corresponding to 
a flux line density $n \sim 5 \cdot 10^{11}\,m^{-2}$. 
After cooling we find that the magnet always levitates near the middle 
between the superconductors, i.e., at a levitation height $h_0 = 
0.5$\,mm.\cite{rf6} In this position the field of the magnet at the surface 
of the superconductors is below $B_{c1}$ and for a quantitative analysis we 
approximate the surface shielding currents induced by the magnet by an image 
dipole model.
In this model the magnetic field on either surface is given by the dipolar 
fields of the magnet and the first image dipole (Fig.\ 2), 
neglecting images further away and also modifications of the field due to 
trapped flux lines. The magnetic moment of the magnet calculated from the 
magnetic remanence of the material and the radius is $M_0 = 4 \cdot 
10^{-6}$\,Am$^2$.
It is simple to calculate the dc sheet current on the surface for a 
particular orientation of the dipole with respect to the surface.\cite{rf7} 

\hspace*{-2.6pt}The dc current distribution 
$\vec{J}=(J_x(x,y,h),J_y(x,y,h))$ 
($x$ and $y$ are coordinates on the surface) is given by $J_x=-2H_y$ and 
$J_y=2H_x$ with
\begin{eqnarray}
H_x &=& \frac{3M_0}{4\pi r^5}(xy\cos\alpha - hy\sin \alpha),\\
H_y &=& \frac{3M_0}{4\pi r^5}\left\{\left( y^2- \frac{r^2}{3}\right)
\cos\alpha - hy\sin \alpha\right\}
\end{eqnarray}
where $r^2 = x^2+y^2+h^2$.
The parallel orientation of the dipole ($\alpha = 0$) has the lowest energy 
but because of trapped flux we cannot rule out that some other orientation 
applies.
Although the variation of the surface current is quite complex and changes 
drastically with the orientation of the dipole\cite{rf7}, the maximum value 
$J_{\mathrm{max}}$ is fairly insensitive to the orientation: 
$J_{\mathrm{max}} = 4400$\,A/m 
when the dipole is perpendicular and $J_{\mathrm{max}} = 5100$\,A/m for the 
parallel orientation at $h=h_0$. In either case the magnetic field is below 
$H_{c1}$, but some flux remains trapped because of the field-cooled 
situation. 

The ac sheet current $\Delta \vec{J}$  due to the oscillations of the magnet 
about its position at $h_0$ is 
\begin{equation}
\Delta \vec{J}(x,y,t) = \left(\left( \frac{\partial J_x}{\partial h}\right) 
_{h_0},\,\left( \frac{\partial J_y}{\partial h}\right) 
_{h_0}\right)
 a \cos(\omega t), \label{gleich1}
\end{equation}
where $a$ is the oscillation amplitude. 
Again the current distribution (shown in Fig.\ 3) is complicated 
but the maximum amplitude of $\Delta J$ is insensitive to the orientation of 
the 
dipole. For an oscillation amplitude $a = 0.5\,\mu\mathrm{m}$ corresponding 
to a
velocity amplitude of $v \approx 1$\,mm/s (at 300\,Hz oscillation 
frequency) a maximum current amplitude of $14$\,A/m is found for the 
perpendicular orientation and $15$\,A/m for the parallel one. The current 
is localized in a small area of the superconducting surfaces below and above 
the sphere, i.e., dissipation occurs only here. 

This ac sheet current is actually an oscillating current density $j$ in the 
samples. Flux lines will be set into motion thereby creating an electric 
field $E = v_f n \Phi _0$, where $v_f$ is the flux line 
velocity. Because $j$ and $E$ are parallel the dissipation per unit volume 
is given by $j \cdot E$. 
In bulk samples the shielding current is flowing 
within the 
London penetration depth 
where it decays exponentially, whereas in thin films the penetration depth 
can exceed the sample thickness. (Note that the Campbell 
penetration depth is an order of magnitude smaller because of the low 
trapped flux density and therefore can be neglected.) 
For simplicity, we use an average current density $\braced{j}= 
\Delta J / \delta$ inside the samples, where $\delta$ is either the 
penetration depth in case of bulk samples or the film thickness. We may then 
describe the dissipation per unit area by $\Delta J \cdot E$, where $E(j)$ 
depends on $\braced{j}$.
\begin{figure}
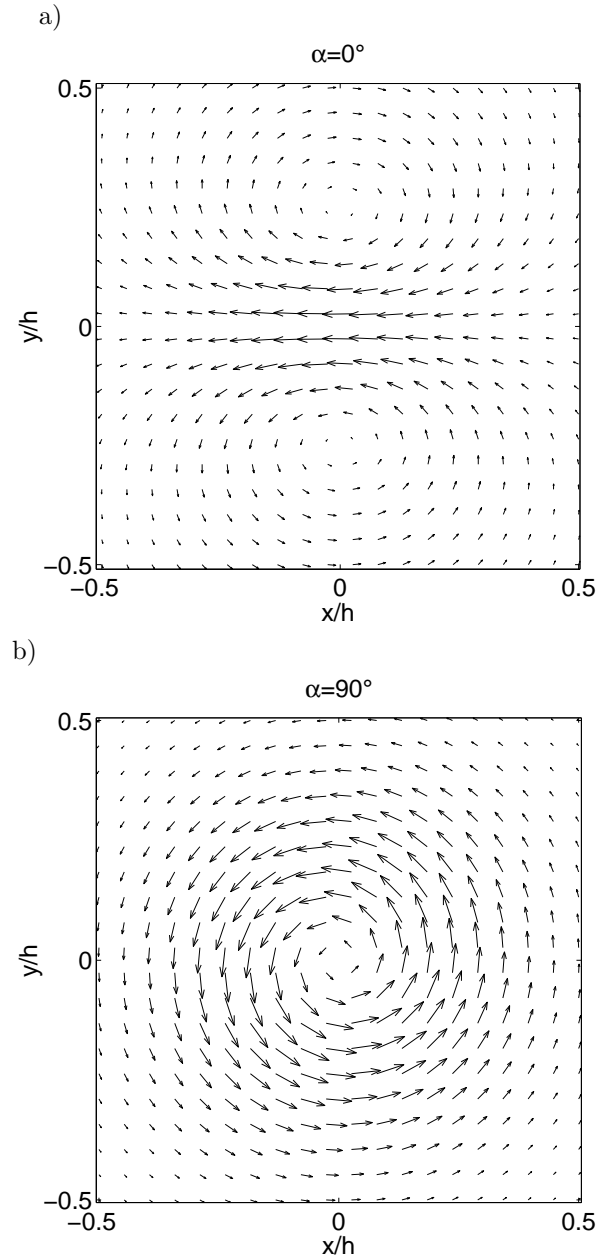

a)\\
{\centering\epsfig{file=figure3a.eps,width=8.5cm}}\\
b)\\
{\centering\epsfig{file=figure3b.eps,width=8.5cm}}\\[1ex]
\begin{minipage}{8.6cm}\caption{Vector plots of the ac surface currents 
calculated from Eq.\ \ref{gleich1}: a) 
for parallel and b) for perpendicular orientation of the dipole. In our case 
$h=0.5\,\mathrm{mm}$ and therefore the displayed surface area is 
$0.5 \times 0.5\,\mbox{mm}^2$ each.
}\end{minipage}
\label{fig43sp}
\end{figure}

We calculate the stationary velocity amplitude of the magnet $v(F)= \omega 
a(F)$ by employing the energy balance between the gain from the driving 
force (which at resonance is in phase with the velocity) and the loss per 
cycle (period $\tau$) on both surfaces:
\begin{equation}
F v \int\limits_{0}^{\tau} \sin ^2(\omega t) \, dt = \frac{F v \pi}{\omega} =
2 \int\limits_{0}^{\tau}\int\limits_{x,y} \Delta J\cdot E\, dx dy dt 
\label{gleich2}
\end{equation}
Because $\Delta J \propto a$ it is evident from Eq.\ 
\ref{gleich2} that a linear dependence of $E(j)$ corresponds to a linear 
response $v(F)$ while a nonlinear response implies a nonlinear 
current-voltage relation. Secondly, it is obvious that the driving force is 
a measure of the electric field, i.e., of the vortex velocity. In the 
following chapter we present our experimental results and analyze them on 
the basis of Eq.\ \ref{gleich2}.
\section{Results and Discussion}

We first describe and discuss the data obtained with the thin film samples 
which are very similar for both the 190\,nm and the 450\,nm thick films.
Then, the measurements on the bulk samples are presented which are very 
similar for both granular and melt-textured YBCO.

\subsection{Thin films}

A summary of our data obtained with the 450\,nm films above 77\,K is shown 
in Fig.~4. At small driving forces we find a steep initial increase of the 
velocity amplitude of the magnet, whereas at driving forces above 
ca.~10\,nN there is only a logarithmic dependence. This implies an 
exponential increase of the dissipation. 
At lower temperatures the velocity amplitudes increase very rapidly with 
drive as shown in Fig.~5 with a largely expanded abscissa. Further increase 
of the driving force amplitude gave unstable oscillations.
Only at very small amplitudes we do find a linear regime between the 
driving force amplitude $F$ and the velocity amplitude $v$, i.e., $F = 
\gamma v$ with a strongly temperature dependent coefficient $\gamma (T)$, 
see Fig.~6. While $\gamma$ decreases considerably below the critical 
temperature $T_c$, it obviously levels off at 4\,K. Here the Q-factor $m 
\omega/\gamma$ which we attribute exclusively to losses in the 
superconductor\cite{em} exceeds $10^6$. The steep drop below $T_c$ follows 
approximately a $(T_c - T)^{-2}$ law as can be seen in the insert of Fig.\ 
6. 

In the linear regime $F=\gamma v$ the induced electric field in the 
superconductors is proportional to the surface current amplitude, $E = R_s 
\Delta 
J$, and the loss (r.h.s. of Eq.~\ref{gleich2}) can be described by a 
surface resistance $R_s \propto \gamma$, namely 
\begin{equation}
R_s = \frac{\gamma \omega ^2 } { S }, 
\label{Rs}
\end{equation}%
where
\begin{equation}
 S  =  2 \int \limits_{x,y} 
\left| 
\left(\left( \frac{\partial J_x}{\partial h}\right) 
_{h_0},\,\left( \frac{\partial J_y}{\partial h}\right) 
_{h_0}\right) 
\right|^2\, dxdy. 
\end{equation}
Here we assume that $R_s$ is a constant on the surface 
of both superconductors. The surface integral $S$ is evaluated numerically. 
In Fig.~6 the $R_s$ values are plotted on the right 
scale for the parallel orientation of the dipole.\cite{error}
\begin{figure}[t]
\centering\epsfig{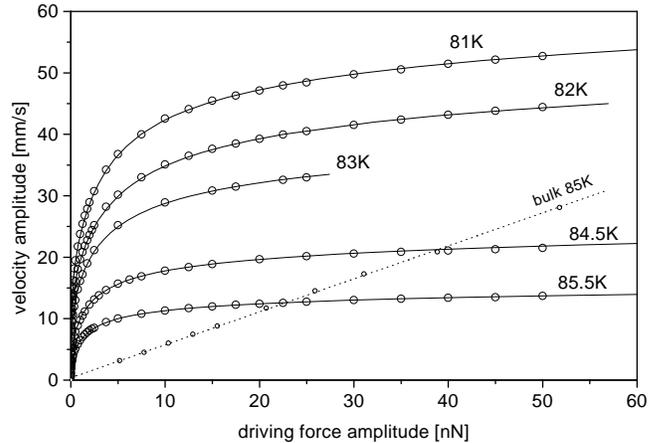}\\[1ex]
\begin{minipage}{8.6cm}\caption{Resonant velocity amplitude of the 
oscillating magnet as a 
function of the driving force at various temperatures. Note the steep 
increase at small driving forces and a logarithmic dependence at large 
drives. The lines are fits of our model to the data, see text. An 
almost linear dependence observed with melt-textured bulk samples 
(from Fig.\ 11) is shown for comparison.}\end{minipage}
\label{fig3}
\end{figure}%
\begin{figure}[b]
\centering\epsfig{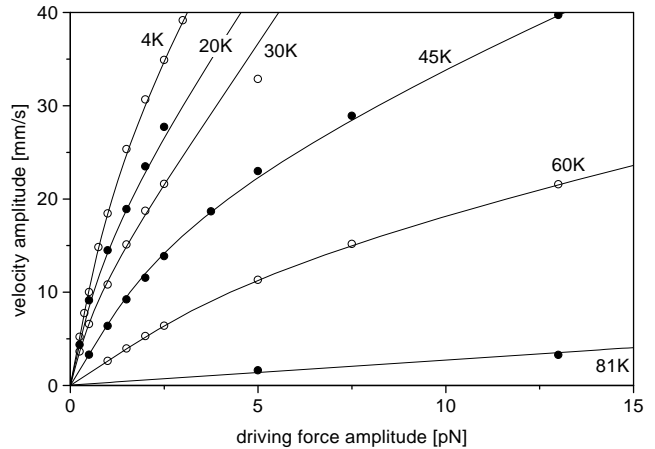}\\[1ex]
\begin{minipage}{8.6cm}\caption{Velocity amplitude at very small driving 
forces 
indicating the linear initial increase at various temperatures. 
The lines are fits of our model to the data, see text.}\end{minipage}
\label{fig2}
\end{figure}

From the surface resistance we can infer the imaginary part of the 
penetration depth of the ac field being given by 
$\lambda''=R_s/\mu_0\omega$.
In our case ($\omega/2\pi \approx 300 ~\mathrm{Hz}$) we find $\lambda''$ to 
drop from 84~nm at 85~K down to only 0.2~nm at 4~K. 
A steep decrease of the surface resistance with temperature has been 
observed at radio and microwave frequencies \cite{belk} and is discussed in 
terms of various models of flux dynamics. In the limit of very low 
frequencies it has been calculated\cite{cc} that the complex resistivity 
$\rho_v$ due to vortex motion is greatly altered when flux creep is 
included. Instead of a purely imaginary resistivity the result has a real 
part given by $\mbox{Re}(\rho_v)=\rho_f \cdot \epsilon$ where 
$\rho_f = B\Phi_0/\eta$ is the usual resistivity for free vortex flow 
($\eta$ is the vortex viscosity) and $\epsilon(\nu) = I_0^{-2}(\nu) \ll 1$, 
where $I_0$ is a modified Bessel function and $\nu = U/2kT$ depends on the 
pinning energy $U(T)$. There is ample evidence\cite{golo} that below 70\,K 
$\nu$ is a temperature independent quantity ranging from 3 to 9. In fact, 
for $\nu=9.5$ our results would agree very well with $R_s = 
{\rm Re}(\rho_v)/t$ ($t$ is the film thickness) if a trapped flux 
density of 1\,mT and typical values for $\eta(T)$ are assumed, see 
Ref.\ \onlinecite{golo} for a compilation of viscosity data. This value for 
$\nu$ appears quite reasonable because, as will be shown below, in the 
nonlinear regime, where $\nu$ begins to drop with increasing current, we 
find $\nu=8$ at a current amplitude of 50\,A/m. It should be mentioned that
recent work on the surface impedance of conventional 
superconductors\cite{placais1} has shown that a more complicated 
electrodynamical description is required. If this theory were directly 
applied to our experiment (thin films, low frequency, low vortex density) 
one would obtain the simple 
relation $R_s \propto B_{c1}^{-2}(T)$ which would agree both in temperature 
dependence (including the region near $T_c$) and in order of magnitude with 
our data. However, this remains to be justified and therefore at present is 
only a speculation.
\begin{figure}[t]
\centering\epsfig{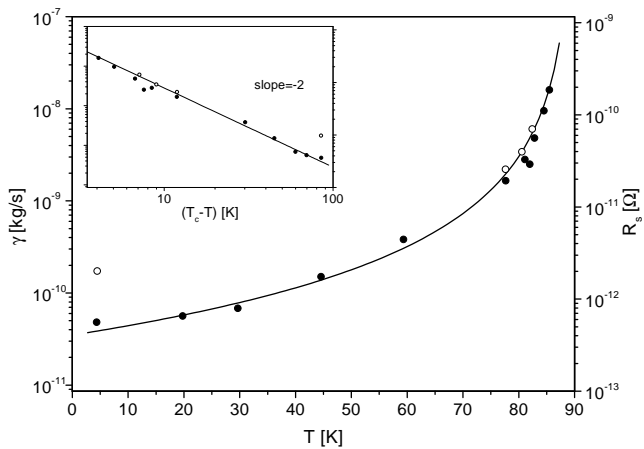}\\[1ex]
\begin{minipage}{8.6cm}\caption{Linear coefficient $\gamma = F / v$ 
from Fig.\ 5 and surface resistance $R_s$ from Eq.\ \ref{Rs}.
Open symbols are data obtained with the thinner film 
$\delta=190\mathrm{nm}$. The solid line is the same as in the insert where
the data are plotted vs $T_c-T$.
}
\end{minipage}
\label{fig1}
\end{figure}

In Fig.~7 the data of Fig.~4 are plotted on a logarithmic scale for the 
driving force $F$. Above 10\,nN we find a logarithmic dependence $v \propto 
\ln (F/F_0)$ which implies an exponential increase of the losses. For 
driving forces between 0.1\,nN and 10\,nN the amplitudes appear to be 
reduced indicating an additional loss mechanism in this regime. Furthermore, 
it is noteworthy that obviously the driving force and 
hence the electric field, or vortex velocity, determines which mechanism 
dominates and not the oscillation amplitude, or shielding current.

In the nonlinear regime of $v(F)$ the electric field in Eq.~\ref{gleich2} 
must have a nonlinear current dependence. This is usually described by 
thermally activated flux motion 
$E(j) \propto \exp(-U(j)/kT)$ with current 
dependent pinning energies $U(j)$, for a review see Ref.\ 
\onlinecite{blatter}.
Essentially  two different dependences are being considered, firstly 
barriers diverging for $j \to 0$:
\begin{equation}
U(j) = U_0 \left( \frac{j_0}{j}\right)^\mu, \label{gleich4}
\end{equation}
where  $0 < \mu < 1$. Vortex glass and collective creep 
models yield this behavior which implies zero dissipation in the limit $j 
\to 0$. Secondly, for large currents, one considers barriers vanishing at a 
critical current density $j_c$ as 
\begin{equation}
U(j)=U_0 \left( 1- \frac{j}{j_c} \right),
\label{gleich5}
\end{equation}
which leads to an exponential dissipation for large currents (Kim-Anderson 
model), $E \propto \sinh(j/j_1)$ with $j_1= j_c kT / U_0$. 
\begin{figure}
\centering\epsfig{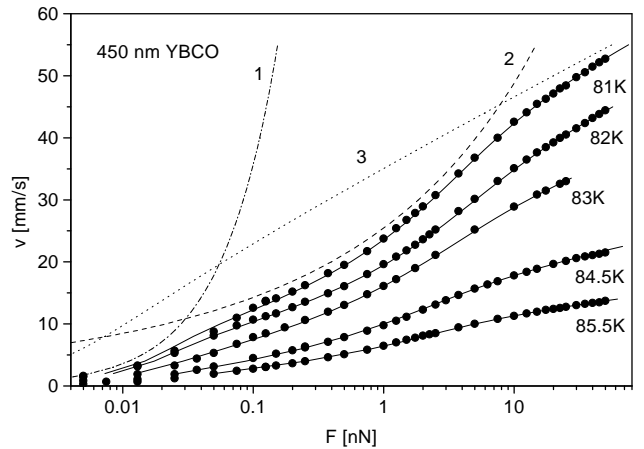}\\[1ex]
\begin{minipage}{8.6cm}\caption{Data of Fig.~4 plotted on a logarithmic 
scale for the driving force. The solid lines are fits of Eq.~\ref{gleich2} 
to the data, see text.
For the 81K curve the contributions of three different mechanisms 
of dissipation are indicated separately: 1 is the linear one (which is an 
exponential 
curve because of the logarithmic abscissa), 2 is the contribution from 
Eq.~\ref{gleich4}, and 3 is from Eq.~\ref{gleich5}.}
\end{minipage}
\label{fig415}
\end{figure}

We fit the 
complete $v(F)$ curves by including both nonlinear mechanisms (Eqs. 
\ref{gleich4} and \ref{gleich5}) and a linear term additively in the 
energy balance, see Eq.~\ref{gleich2}. An example is shown in Fig.~7, 
where the separate contributions and their sum are compared with an 
experimental curve.
The linear term (surface resistance) contributes only at very low 
amplitudes. In the intermediate 
range the glass (or creep) term (Eq.~\ref{gleich4} with $\mu=0.2$) is 
dominant. Only at the largest drives the Kim-Anderson term 
(Eq.~\ref{gleich5}) becomes relevant. It is this mechanism which leads to 
the observed slow logarithmic increase of the oscillation amplitudes. 
From these fits we can determine the barriers $U(j)$ at various 
temperatures, see Fig.~8. It is a peculiarity of Eq.~\ref{gleich5} that 
only the constant slope $-dU/dj=U_0/j_c$ can be obtained reliably from a 
fit to the data because the quantity $U_0$ determines only the prefactor 
which we do not evaluate for it contains other unknown quantities, e.g. the 
density of the trapped flux $n$ or the prefactor of the vortex velocity 
$v_f$. From Fig.~8 it is evident that at large currents and high 
temperatures the barriers of Eq.~\ref{gleich5} ultimately drop below 
those of Eq.~\ref{gleich4}. 
For the exponent $\mu$ determining the barriers we find $\mu =0.2$ 
which is close to the theoretical value 1/7 for single vortex pinning 
which is resonable for the small vortex densities in our 
experiment.\cite{rf10} We should mention that the exponential 
fitting parameters are not affected by the particular current distribution 
used (parallel or perpendicular orientation of the dipole), only prefactors 
are different because the parallel dipole leads to a lower total dissipation 
than the perpendicular one.
\begin{figure}
\centering\epsfig{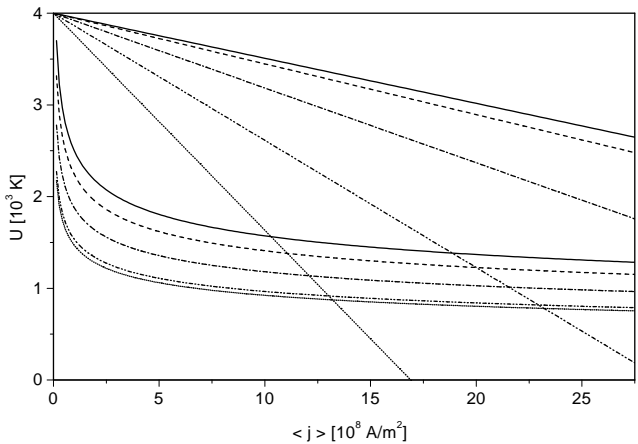}\\[1ex]
\begin{minipage}{8.6cm}\caption{Dependence of the pinning potential $U$ on 
the average current density in 
the film obtained from the fits of our model to the data: curved lines are 
Eq.~\ref{gleich4}, straight lines are Eq.~\ref{gleich5} at the following 
temperatures (from top to bottom): 78~K, 81~K, 83~K, 84.5~K, 85.5~K.
Note that from Eq.~\ref{gleich5} we can determine only the slopes $U_0/j_c$; 
in the figure we arbitrarily set $U_0=4 \cdot 10^3 \mbox{K}$
}\end{minipage}
\label{fig5}
\end{figure}

It is interesting to note that all three mechanisms of dissipation are  
found to coexist at the same temperature but dominate in different regimes 
of the driving force. Furthermore, we find that the current 
$j_1$ which sets the scale for the exponential dissipation extrapolates to 
zero at 86.5~K (well below $T_c=89.5$~K) which might indicate an onset of 
free flux flow where the magnet loses its lateral stability.

Towards lower temperatures only the linear term and the glass term 
contribute as no data could be obtained at larger drives because the 
oscillations became unstable. For temperatures below 60\,K we find that the 
evaluated barriers decrease nearly linearly with temperature, see 
Fig.~9. We find $U=16T$ and $U=10T$ (i.e., $\nu=8$ and 5) for current 
amplitudes of 50\,A/m and 500\,A/m, respectively. The ratio 1.6 is given by 
the factor $10^{0.2}$, see Eq.\ \ref{gleich4}. The linearity of $U(T)$ is 
again a signature of flux creep, but now with a current dependent creep 
factor. The values for $\nu$ are determined by the logarithm of various 
properties of the pinning centers and of the inverse frequency. For this 
reason our values obtained at 300\,Hz are larger than those obtained at 
microwave frequencies\cite{belk} but lower than those obtained from magnetic 
relaxation measurements.
Approaching $T_c$ the effective barrier finally decreases (see Fig.\  
8) because the quantity $j_0$ in Eq.~\ref{gleich4} which is related to the 
critical current density\cite{blatter} goes to zero.
\begin{figure}
\centering\epsfig{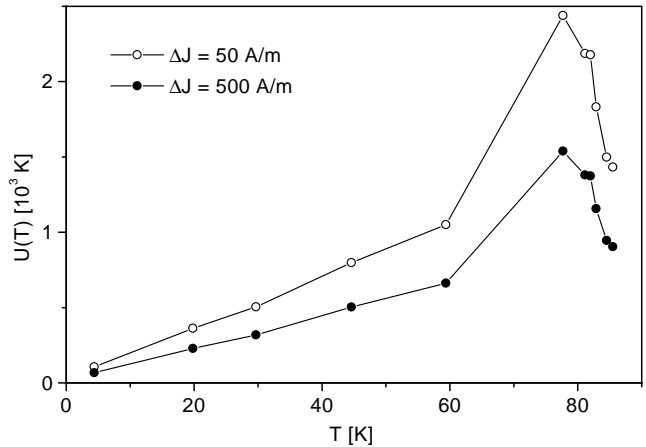}\\[1ex]
\begin{minipage}{8.6cm}\caption{Temperature dependence of the effective 
barriers (Eq.~\ref{gleich4}) for two different current 
amplitudes. Below 60\,K the dependence is linear for both currents with 
slopes 10 and 16, respectively. The solid lines are merely a guide to the 
eye.
}\end{minipage}
\label{fig420}
\end{figure}

\subsection{Bulk samples}
We will now give an overview of the measurements with YBCO bulk 
samples and compare these results with the thin films. 
We have found that the dissipative behavior of melt-textured and sintered 
bulk samples is quite similar. Measurements on the sintered samples have 
already been published\cite{apl1} and therefore we will focus on our 
melt-textured results. We emphasize, however, that our present 
interpretation of the data is more complete and to some extent makes our 
earlier analysis obsolete.

Figures~10 and 11 show the velocity amplitude of the magnet as a 
function of driving force at different temperatures for melt-textured 
and sintered samples. At 4\,K we find a nonlinear increase of the amplitude 
even at small amplitudes followed by a nearly linear dependence for large 
drives. This initial nonlinear increase is less pronounced at higher 
temperatures where we find approximately linear curves up to 85\,K. For 
higher temperatures additional dissipation arises and above 88\,K 
the dependence is simply $v \propto \sqrt{F}$ indicating a quadratic 
damping, i.e., losses proportional to the cube of the amplitude.
Because of the various mechanisms of dissipation (to be discussed 
below) dominating at diffent temperatures and amplitudes, there exists a 
temperature range below $T_c$ where the oscillation amplitudes 
surprisingly increase with temperature at constant drive, corresponding to a 
minimum of dissipation. 
This can be seen more clearly in Fig.~12 where the velocity amplitudes are 
shown as a function of temperature for various constant drives. We note that 
for sintered YBCO the maximum is much more pronounced, see 
Ref.~\onlinecite{apl1}. 
\begin{figure}
\centering\epsfig{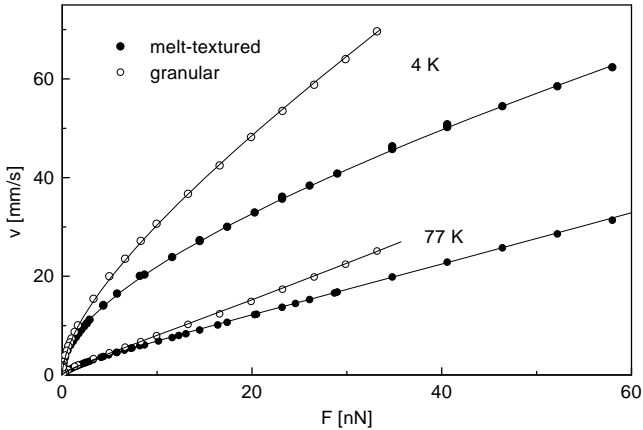}\\[1ex]
\begin{minipage}{8.6cm}\caption{Velocity amplitudes as a function of the 
driving force for sintered and melt-textured samples at 4\,K and 77\,K. The 
lines are fits of our model to the data.
}\end{minipage}
\label{fig10}
\end{figure}%
\begin{figure}
\centering\epsfig{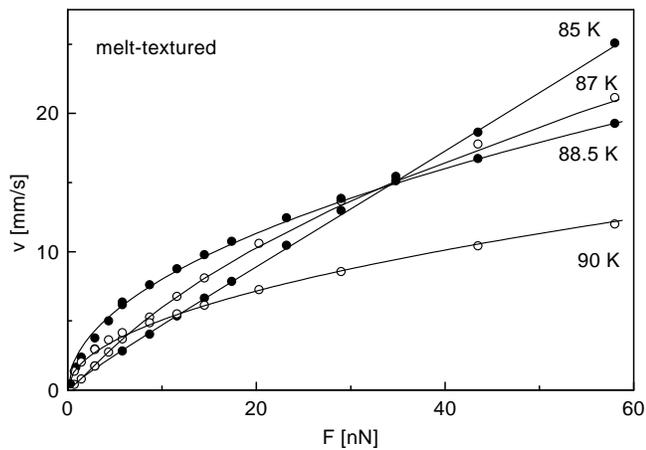}\\[1ex]
\begin{minipage}{8.6cm}
\caption{Velocity amplitudes versus driving force at temperatures above 
85\,K where damping becomes quadratic leading to a square-root
dependence. The lines are fits of our model to the data.}
\end{minipage}
\label{fig46mt}
\end{figure}
Considering the initial nonlinear increase of the amplitude at 4\,K
followed by a nearly linear rise for larger drives we find that this 
behavior can be well fitted by adding a linear dissipation and 
the collective flux creep (Eq.~\ref{gleich4}), using an exponent $\mu = 
0.14$. 
Because Eq.~\ref{gleich4} describes the data over almost the 
entire range the exponent $\mu$ can be determined with much better accuracy 
than with the thin-film samples: increasing $\mu$ to 0.2, e.g., leads to a 
clearly poorer fit.
From the linear term (dominating only
at the smallest amplitudes) we 
infer a surface resistance which is of the order of $R_s \sim 
10^{-10}\,\Omega$ at 4\,K rising to $10^{-8}\,\Omega$ at 85\,K.
Comparing these results with our data on YBCO thin films we note that at 
78\,K $R_s$ has increased by three orders of magnitude.

At 4\,K the barrier in the nonlinear regime is much smaller than at 
high temperatures (there are no experimental data between 4\,K and 77\,K). 
Above 77\,K the barriers decrease with increasing temperature. 
These results are similar to those obtained with the films except for the 
fact that the barriers in the bulk are a factor of 2 smaller than in the 
films.

\begin{figure}
\centering\epsfig{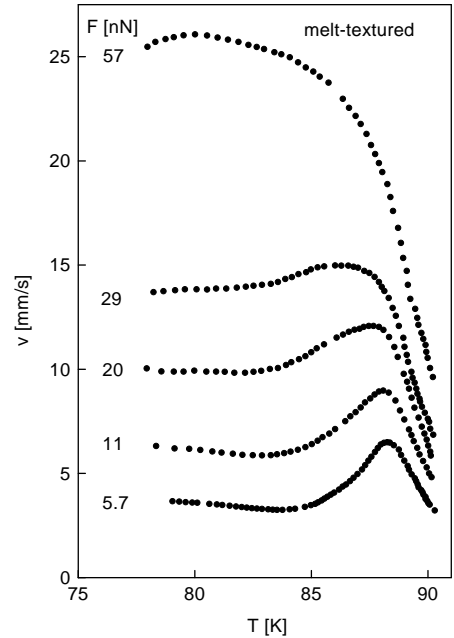}\\[1ex]
\begin{minipage}{8.6cm}\caption{Temperature dependence of the velocity 
amplitude at various constant driving forces. Note the maximum at about 
88\,K (i.e., minimum of dissipation).
}\end{minipage}
\label{fig47mt}
\end{figure}
For temperatures above 85\,K quadratic damping rapidly increases and 
eventually becomes to be the only relevant dissipation above 88\,K. We 
attribute this to a mechanism proposed by Melville\cite{melville} for 
conventional superconductors in which single vortices are considered being 
trapped perpendicular to the surface of the superconductor. When an external 
field smaller than $H_{c1}$ is applied shielding currents are induced at 
the surface producing a Lorentz force on the trapped vortices. If this force 
is strong enough to depin the vortices they will bend in the direction of 
the applied force, while they stay pinned in the interior of the bulk where 
no current is flowing.
The amount of bending is given by the equilibrium between the Lorentz 
force and the elastic force of the vortex. When the applied field is 
reduced and finally reversed the vortex moves back in an irreversible way, 
so that the flux component parallel to the surface passes through a 
hysteresis. This leads to dissipation where the loss per cycle and surface 
area is given by $B\,H^3 / H_{c1}\,j_c$,
where $H$ is the amplitude of the external field and $j_c$ the depinning 
current density. In our experiment, $H\propto a$ and therefore, with 
Eq.\ref{gleich2}, this mechanism leads to quadratic damping $F 
\propto v^2$ having a coefficient which diverges at $T_c$ as was found 
experimentally already for sintered samples.\cite{apl1}

It should be noted that the well known Bean critical state also leads to 
quadratic damping and is used successfully in the interpretation of most 
other levitation experiments involving larger magnets. However, in our 
experiment the magnetic field is low and a critical state is not 
reached. Also, in the model of Melville the dissipation is proportional to 
the density of trapped flux $B$ which makes it plausible that the losses in 
the melt-textured samples are larger than those in the sintered samples.

Besides the much larger damping found in bulk samples, there are two main 
features distinguishing the bulk results from the thin film ones: Firstly, 
for bulk samples the damping becomes quadratic approaching $T_c$ so that 
the amplitudes exceed those for thin films (Fig.~4) where the 
dissipation grows exponentially with amplitude. We believe that the reason 
for this is the sample thickness. In bulk samples the vortices stay
pinned in the interior and move just within the penetration depth. In thin 
films the vortices are exposed to the driving current over their entire 
length. This leads to high flow rates of the vortices in the films whereas 
in bulk samples the flow rates are limited due to pinning deeper inside the 
samples.

Secondly, for bulk samples there is a minimum in dissipation below $T_c$ 
while for thin films the damping grows monotonically with temperature. This 
minimum might be due to a softening of the flux line lattice leading to 
improved pinning because the vortices can adjust more effectively to a 
state of lower total energy. At still higher temperatures pinning becomes 
less effective, hysteretic damping occurs and dissipation diverges 
approaching $T_c$. In thin films the exponential growth of dissipation 
dominates and no minimum in damping occurs. 

\section{conclusion}
In summary, the damping of the oscillations of the levitating magnet is due 
to energy losses caused by flux motion of various dynamic regimes. At 
sufficiently small amplitudes linear friction is observed which can be 
described by a surface impedance having particularly low values for thin 
films at low temperatures. The temperature dependence of the surface 
impedance is discussed in terms of a vortex resistivity caused by flux 
creep. Towards larger amplitudes we find losses to originate from the motion 
of collectively pinned vortices. From our analysis pinning barriers are 
obtained having the typical inverse current dependence. The effective 
barriers increase linearly with temperature up to 60\,K as expected from 
creep effects. Finally, at even larger amplitudes, the losses in the films 
grow exponentially and eventually exceed those in the bulk samples. This 
seems plausible because in the films the flux lines are exposed to large 
current densities along their entire length, whereas in the bulk the Lorentz 
forces decay with increasing distance from the surface.

When comparing our experimental method (which was motivated by an interest 
in the physics of superconducting levitation) with more standard techniques 
for investigating vortex dynamics in superconductors, like measurements of 
the ac susceptibility, of the current voltage characteristics ,of the 
magnetic relaxation, or mechanical oscillators containing superconducting 
samples, we note several obvious differences some of which are 
disadvantageous while others are of advantage. Firstly, our method is not 
affected by demagnetization factors or edge effects, the geometry is just a 
magnetic dipole in the middle between essentially infinite and flat 
superconducting surfaces. Secondly, it is a resonance method with high 
sensitivity and wide dynamic range: The dissipated power varies from 
$10^{-9}$\,W down to $10^{-17}$\,W and the electric fields caused by vortex 
motion range from $10^{-4}$\,V/m down to $10^{-11}$\,V/m. Because of the 
resonance method, though, the frequency cannot be varied. The magnetic field 
is inhomogeneous and rather small, and the distribution of trapped flux 
lines is unknown. Nevertheless, the damping of the oscillating magnet can 
be described quantitatively by various mechanisms of flux dynamics with flux 
creep being an important issue.

\section*{acknowledgments}
We are grateful to K.F.\ Renk for support of our work, to E.V.\ Pechen 
for providing the films and to H.J.\ Bornemann for the melt-textured 
material. We had helpful discussions 
with N.\ L\"utke-Entrup and B.\ Pla\c{c}ais on the interpretation of 
surface impedance measurements.
Financial support by the Deutsche Forschungsgemeinschaft is 
gratefully acknowledged by R.G. and M.N. (Graduiertenkolleg ``Komplexit\"at 
in Festk\"orpern'').

\section*{APPENDIX: elastic forces on the oscillating microsphere}
The elastic forces on the levitating microsphere were investigated  by 
measuring the amplitude dependence of the resonance frequency $\omega(a)$ 
and also of the second and third harmonic amplitude. In addition, the effect 
of a 
large dc voltage (up to $\pm 1000\,\mathrm{V}$) across the capacitor on the 
above quantities was determined. By and large, the oscillator displays 
typical nonlinear properties.\cite{mook} The ``backbone curve'' $\omega 
(a)$, e.g., is the usual parabola for the thin-film samples below 81\,K. 
However, at higher temperatures and for all bulk samples in particular, 
stronger dependences are found which indicate that a third order Taylor 
expansion of the return forces at the equilibrium position is no longer 
adequate. On the other hand, the coefficients of the second and third 
harmonic are found to be independent of temperature. Application of the dc 
voltage changes the equilibrium position by few tens of micrometers as can 
be estimated from the stiffness. Furthermore, the resonance frequency 
changes because of the changing tension, in analogy with a vibrating string: 
$\omega 
^2$ is proportional to the static force as observed earlier with a niobium 
capacitor.\cite{niob} The coefficient of the second harmonic is strongly 
affected and can be reduced to zero at a particular voltage, which implies a 
symmetric return force at the new equilibrium position.

These elastic properties may be discussed in terms of the 
image-dipole model presented in Ch.\ III. Neglecting higher order images the 
resonance frequency is easily calculated to be $\omega/2\pi\approx 
200\,\mathrm{Hz}$ if the equilibrium position is in the middle between the 
two surfaces. Gravitation and especially trapped flux lead to unsymmetric 
forces and therefore to an unsymmetric position and hence to a second 
harmonic amplitude. The force between a single 
vortex and a magnetic dipole perpendicular above a superconducting thin film 
has been considered in Ref.\ \onlinecite{china}. Because we have no 
quantitative 
information on the distribution and the orientation of the trapped vortex 
lines in the samples we cannot calculate their contribution to the elastic 
forces on the levitating magnet. Therefore, the image-dipole model remains a 
rather crude description of the elastic forces.
For lateral stability of the magnet, which is a prerequisite of our 
method, trapped flux is clearly necessary.

\end{multicols} 
\end {document}